\documentclass{optica-article}

\journal{opticajournal} 
\usepackage{graphicx} 
\usepackage{dcolumn} 
\usepackage{bm} 
\usepackage{color}
\usepackage{float}
\usepackage{dsfont}
\usepackage{subfig}
\usepackage{hyperref}
\usepackage{listings}
\usepackage{braket}
\usepackage{siunitx} 
\articletype{Research Article}

\newcommand{\PrYSO}{Pr$^{3+}$:Y$_2$Si{O$_5\,$}}
\newcommand{\maxeff}{\qty[uncertainty-mode=separate]{62\pm2}{\percent}}
\newcommand{\maxeffround}{\qty{62}{\percent}}
\newcommand{\afclimit}{\qty{54}{\percent}}

\begin{document}
	
	\title{Efficient cavity-assisted storage of photonic qubits in a solid-state quantum memory}
	
	\author{Stefano Duranti \authormark{1}, S\"oren Wengerowsky\authormark{1,*}, Leo Feldmann\authormark{1}, Alessandro Seri\authormark{1}, Bernardo Casabone\authormark{1} and Hugues de Riedmatten\authormark{1,2,**}}
	
	\address{\authormark{1}ICFO-Institut de Ciencies Fotoniques, The Barcelona Institute of Technology, Mediterranean Technology Park, 08860 Castelldefels (Barcelona), Spain}
	\address{\authormark{2}ICREA-Instituci\'{o} Catalana de Recerca i Estudis Avan\c cats, 08015 Barcelona, Spain}
	
	\email{\authormark{*}soeren.wengerowsky@icfo.eu,\authormark{**} hugues.deriedmatten@icfo.eu }

	\begin{abstract*} 
		We report on the high-efficiency storage and retrieval of weak coherent optical pulses and photonic qubits in a cavity-enhanced solid-state quantum memory. By using an atomic frequency comb (AFC) memory in a \PrYSO crystal embedded in an impedance-matched cavity, we stored  weak coherent pulses at the single photon level with up to {\maxeffround} efficiency for a pre-determined storage time of \qty{2}{\us}. We also confirmed that the impedance-matched cavity enhances the efficiency for longer storage times up to \qty{70}{\micro\s}. Harnessing the temporal multimodality of the AFC scheme, we stored weak coherent time-bin qubits with a record \qty[uncertainty-mode=separate]{51\pm2}{\percent} efficiency and a fidelity over \qty[uncertainty-mode=separate]{94.8\pm1.4}{\percent}, limited by imperfections in the qubits creation and measurement.
		
	\end{abstract*}

	\section{Introduction}
	Photonic quantum memories are devices that can store quantum information carried by photons and retrieve it at a later time \cite{afzelius2015quantum}. They can be used as synchronization devices for probabilistic quantum processes and, as such, find applications in quantum light engineering, photonic quantum computation and quantum networking. In particular, they are crucial elements for the implementation of a large scale quantum internet \cite{kimble2008quantum} by enabling long-distance distribution of entangled states using quantum repeaters \cite{briegel1998quantum, sangouard2011quantum}.
	
	Quantum memories have been implemented in various ensemble-based systems such as cold atomic clouds \cite{Chaneliere2005,chou2005measurementinduced,cho2016highly,cao2020efficient,wang2021cavityenhanced}, warm atomic vapours \cite{julsgaard2004experimental,reim2011singlephotonlevel,hosseini2011unconditional,Kaczmarek2018,Buser2022,Wang2022}, nanomechanical resonators \cite{Wallucks2020} and rare-earth doped solids. Single quantum systems such as single atoms \cite{Specht2011,korber2018decoherenceprotected}, vacancy centers in diamonds \cite{Humphreys2018,Bhaskar2020,Pompili2021} and  single trapped ions \cite{Stute2012,wang2017singlequbit,Stephenson2020,Krutyanskiy2022} have also been used as quantum memories. Important progress has been realized in recent years on quantum memory performances. Storage and retrieval efficiencies close to \qty{90}{\percent} have been demonstrated in high optical depth cold atomic gases \cite{cao2020efficient,cho2016highly,wang2019efficient}, as well as a storage time of several hundreds of ms \cite{yang2016efficient,wang2021cavityenhanced} in separate experiments.

	Rare-earth-ion doped crystals are interesting for quantum memory applications because they enable ensemble-based storage without the need for complex laser cooling and trapping schemes. They provide a large number of atoms with optical and spin transitions that are naturally trapped in a solid-state matrix. The ions exhibit long coherence times at cryogenic temperature, both for the optical and spin transitions, which have led to demonstrations extremely long storage times for classical light \cite{heinze2013stopped, ma2021onehour,holzapfel2020optical}. The highest efficiency achieved so far for the storage of weak coherent pulses at the single photon level was \qty{69}{\percent} using the gradient-echo-memory protocol \cite{hedges2010efficient}. This demonstration did not include the storage of qubits. Although the efficiencies achieved with rare-earth-ion doped crystals are still currently lower than in cold atomic gases, they possess a large multiplexing capacity. Their inhomogeneous broadening over several GHz, which can be tailored by spectral hole burning techniques,  is a resource for temporal and spectral multiplexing \cite{ortu2022multimode}. In particular, the atomic frequency comb (AFC) protocol~\cite{afzelius2009multimode}, based on a comb-shaped structure in the inhomogeneous broadening, exhibits intrinsic temporal multi-modality~\cite{lago-rivera2021telecomheralded,businger2022nonclassical}. It has also proven to support  frequency~\cite{sinclair2014spectral,seri2019quantum} and spatial~\cite{zhou2015quantum,yang2018multiplexed} multiplexing.
	
	The AFC protocol, first demonstrated with weak coherent qubits in 2008~\cite{deriedmatten2008solidstate}, plays an important role in rare-earth based quantum memories, as it is the only protocol so far that has allowed the storage of externally generated single and entangled photons in such devices. This enabled the demonstration of quantum repeater building blocks, e.g. quantum entanglement between quantum memories and telecom photons \cite{clausen2011quantum, saglamyurek2011broadband,rakonjac2021entanglement}, light-matter quantum teleportation \cite{Bussieres2014,LagoRivera2023} or heralded entanglement between spatially separated quantum memories \cite{lago-rivera2021telecomheralded,Liu2021}. Moreover, the AFC intrinsic temporal multiplexing capability allows a significant speed-up in entanglement distribution rate \cite{simon2007quantum,lago-rivera2021telecomheralded,businger2022nonclassical} in quantum networks as well as a natural platform for the storage of time-bin encoded states.

	However, reaching high efficiencies for AFC memories requires a large optical depth which is challenging to achieve with mm or cm sized doped crystals. The highest efficiency demonstrated so far in free-space crystals is around \qty{30}{\percent}  for single photon qubits \cite{maring2018quantum}, \qty{38}{\percent} for weak coherent states~\cite{horvath_noise-free_2021} and \qty{40}{\percent} for classical light \cite{ortu2022multimode}. Moreover, the AFC efficiency, is typically limited by re-absorption to {\afclimit} \cite{afzelius2009multimode} for retrieval in the forward direction. A possibility to surpass this limit is provided by cavity-assisted AFC, where the crystal is embedded in a low finesse cavity in the impedance-matched regime \cite{afzelius2010impedancematched}.   
	In the last decades, { several experiments employing this technique have achieved remarkable efficiencies of up to \qty{56}{\percent} with classical pulses~\cite{sabooni2013efficient,jobez2014cavityenhanced}, \qty{27.5}{\percent} at the single photon level and \qty{7}{\percent} for optical qubits using an integrated design  ~\cite{davidson2020improved}. Achieving high efficiencies at the single photon level presents additional experimental challenges in terms of cavity stability and noise suppression.}    

	 In this paper, we demonstrate an impedance-matched quantum memory at the single photon level. We report the highest AFC efficiency achieved so far with up to {\maxeffround} for weak coherent pulses as well as the highest storage and retrieval efficiency (up to \qty{51}{\percent}) for time-bin qubits in a solid-state device.
	
	\section{Theoretical Background}
	
	For an AFC memory in a low-finesse cavity, the impedance matching condition is given by \cite{afzelius2010impedancematched}: 
	\begin{equation}
		\label{eq:imcondition}
		R_\text{in}=R_\text{out} e^{-2\tilde{d}}
	\end{equation}
	Here, R$_\text{in}$ is the reflectivity of the coupling mirror, R$_\text{out}$ the reflectivity of the second mirror, $\tilde{d}$ is the effective optical depth of ions averaged over the comb peaks. In the case of an AFC with square peaks, this last quantity can be evaluated as $\tilde{d}=\frac{d}{\mathcal{F}_\text{AFC}}$ \cite{afzelius2009multimode}, where $d$ is  the optical depth of the crystal and $F_{AFC}$ is the finesse of the comb, namely the ratio between the distance among the peaks and their width. If Eq.~\ref{eq:imcondition} is fulfilled, for perfect mode matching  and if there is no intracavity loss, the input light is coupled to the cavity and fully absorbed in the crystal. In that case, the efficiency of the AFC is given by \cite{afzelius2010impedancematched}: 
	\begin{equation}
		\label{eq:im_eff}
		\eta_\text{cav}=\frac{4\tilde{d}^2 e^{-2\tilde{d}} (1-R_\text{out}e^{-2\tilde{d}})^2 R_\text{out} \eta_\text{deph}}{(1-R_\text{out} e^{-2\tilde{d}})^4}
	\end{equation}
	where $\eta_\text{deph}$ depends on $F_{AFC}$. $\eta_\text{cav}$ can in principle reach \qty{100}{\percent} even for excited state storage, if $F_{AFC}\gg1$, $ R_\text{out}=1$ , $ \tilde{d}\ll 1 $ and for negligible intra-cavity loss. In the presence of an intra-cavity round-trip loss $\epsilon$, $R_{out}$ can be replaced by $R_{out}-\epsilon$ in Eq.~\ref{eq:im_eff} to model a realistic experiment~\cite{jobez2014cavityenhanced}. For a given intra-cavity loss, there is an optimum average optical density $\tilde{d}$ for which the efficiency is maximal.
	
	\section{Experimental Setup}

	\begin{figure}[hbtp]
		\centering
		\includegraphics[scale=0.6]{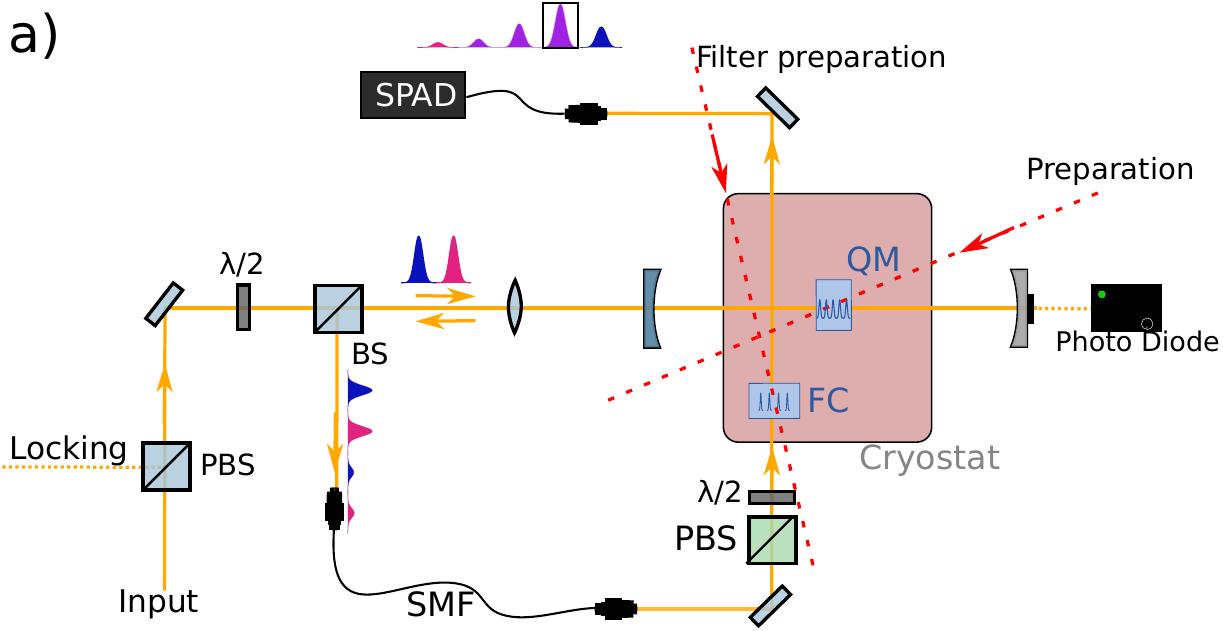}
		
		\vspace{2ex}
		
		\includegraphics[scale=0.45]
		{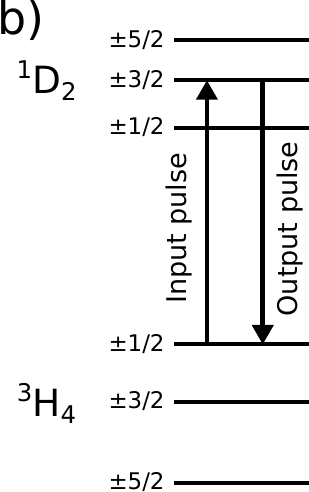}\hspace{3ex}\includegraphics[scale=0.25]
		{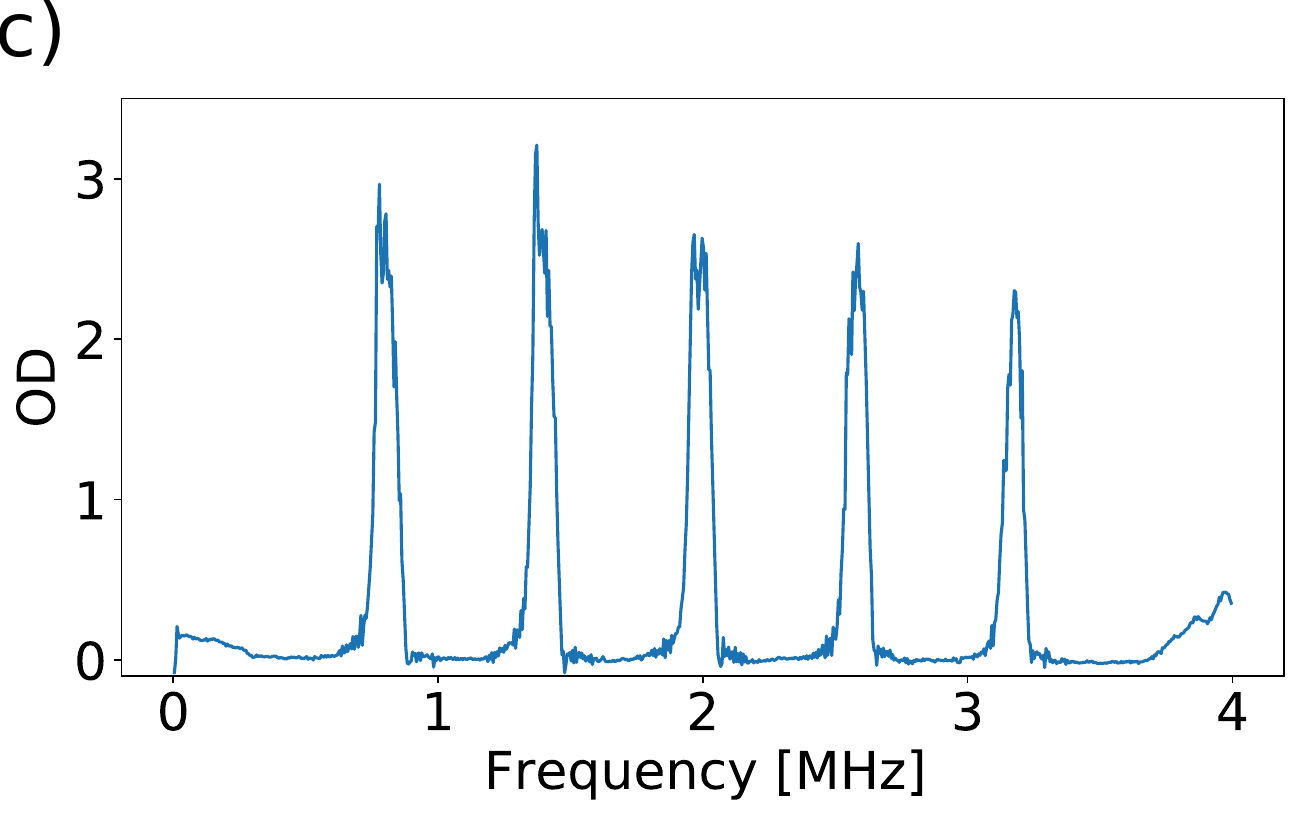}
		\caption{\raggedright (a): Sketch of our experimental setup for qubit storage. The two pulses constituting the time-bin qubit are depicted as { red} and blue. The slightly { smaller} pulses are the echoes, preceded by residual reflections from the cavity due to imperfect mode and impedance matching. { At the detector, constructive interference is depicted with the time bin of interest shown by a thin box. } (b): Level scheme of the  ${}^3H_4-{}^1D_2$ transition of \PrYSO shows 3 hyperfine levels in both the ground and excited state. The optical transition, where the AFC is prepared, is  $\pm1/2g$ -- $\pm3/2e$. (c): Plot of a single pass absorption measurement of the AFC. The finesse of the AFC is 5.8.} 
		
		\label{fig:setup_pic}
		
	\end{figure}
	
	\bigskip 
	
	In this experiment, we use two \PrYSO  crystals, one functioning as a memory crystal and the other one as time-bin qubit analyzer using an AFC-based interferometer {for qubit analysis. For measurements of the AFC efficiency, the beam path leading to the second crystal was not used, as the fiber patch cable was directly connected to the detector.} 
	The  ${}^3H_4-{}^1D_2$ - transition used is at \qty{605.977}{\nano\meter}, in the visible regime (orange light). This transition is shown in the bottom part of Fig.~\ref{fig:setup_pic}.
	Both crystals are cooled down to \qty{3.2}{\kelvin} in a closed-loop Montana cryostat.
	Fig.~\ref{fig:setup_pic} shows a schematic of the optical setup used. 
	The laser to prepare the spectral structure and input pulses, is a frequency-doubled external cavity diode laser at \qty{606}{\nm}. The fundamental wavelength at \qty{1212}{\nm} is locked to an ultra-stable cavity to achieve a line width below \qty{3}{\kHz} { (FWHM)}. Acousto-optic modulators (AOMs) are used in double-pass to create the pulses used to prepare the AFCs, to lock the cavity and to be stored in the memory.

	By means of a PBS, two different beams of light, the signal beam and the locking beam,  are overlapped and  coupled into the cavity. The locking beam is shifted by \qty{640}{\mega\Hz} and has orthogonal polarization with respect to the signal beam, in order to minimize interactions with the ions. A half-wave plate (HWP) is then used to orient the signal beam polarization along the interacting axis of the crystal. Then, { after passing through a 50:50 beam-splitter (BS), the beam reaches} the cavity. The highly asymetric cavity is formed by two mirrors with reflectivities $R_\text{in}$=0.4 and $R_\text{out}$=0.97 { and curvatures of \qty{125}{\mm} and \qty{100}{\mm}, respectively,} leading to a finesse $F_{cav}$ = 6.5. The cavity { of length \qty{208}{\mm} is built around the vacuum chamber such that the \PrYSO-crystal is located at its waist.} We pick up half of the reflected light by means of the BS, and we couple it to a single-mode fiber. Depending on whether we store qubits or not, this light is then out-coupled in free-space and sent to the analyzing  crystal or directly detected in a single-photon avalanche detector (SPAD). After crossing the analyzing crystal, the light is coupled into a single-mode fiber and sent to the SPAD. The locking light, transmitted through the cavity is detected using a photo diode, and read by an Arduino microcontroller. This microcontroller { uses a simple algorithm to maximize the transmission through the cavity and } feeds back on the voltage applied to the piezo that controls the cavity length. { Therefore it keeps the cavity on resonance with the locking beam during the locking period. }

	The preparation beam used to create AFCs in the memory crystal is sent on another path, crossing the signal beam inside the crystal with an angle of \qty{7}{\degree} (not coupled to the cavity mode). We prepare AFCs using spectral hole burning techniques described in~\cite{jobez2016highly}. The AFC is prepared with a target average OD  of $\tilde{d}=\frac{1}{2} \ln(\frac{\text{R}_\text{out}}{\text{R}_\text{in}})=0.46$ which ensures impedance-matching according to Eq.~\ref{eq:imcondition}. 	  
	Analogously, a strong preparation beam addresses the filter crystal, in order to prepare an AFC where the transmitted pulses have the same amplitude as the stored and retrieved pulses to form an unbalanced Mach-Zehnder interferometer for qubit analysis.

	\section{Experimental Procedure}  
	A new AFC is prepared at every cycle of the cryostat, i.e. every second.	
	Each experimental cycle is divided into two phases: (1) a preparation and locking phase lasting around \qty{160}{\milli\s}, during which the cavity is stabilized and the comb structure is prepared. During this phase, all the shutters on the detector path are closed to avoid strong light to reach the APDs. (2) A measurement phase, lasting up to \qty{80}{\milli \s}, when pulses to be stored are sent to the quantum memory. In the second phase, the locking light shutter and preparation shutter close, and only the single photon light is entering the cavity path. { The locking algorithm stops and a constant voltage offset is applied to the last position of the cavity in order to be on resonance with the input photons.} To emulate single photons, we use weak coherent input states containing  on average $\mu_{in} < 1 $ photons per pulse, in front of the cavity. We send 1000 pulses separated by a time ranging from \qty{13}{\micro\s} to \qty{80}{\micro\s}, depending on the AFC storage time $\tau _{AFC}$. Given the unbalanced reflectivity of the cavity mirrors, the AFC echoes are mostly leaving the cavity through the input mirror. We detect those echoes in the same path as the input pulses.

	To account for instabilities in laser power or alignment, every other cryostat cycle we prepare a high OD absorption feature instead of an AFC inside the crystal.{  The OD of this feature about 3.5 and the width is \qty{1.6}{\MHz}.} This effectively blocks the cavity and, as a consequence, we only detect a signal corresponding to \qty{40}{\percent} of the input light, that gets reflected from the first mirror.
	
	The storage efficiency is estimated by comparing these reflected pulses (normalized to \qty{100}{\percent} to have an estimation of the input pulse) with the AFC echoes retrieved from the quantum memory.  
	To ensure that the AFC efficiency is not overestimated, the coupling to the SM-fiber is optimized for the input mode, i.e. the spatial mode that is reflected from the first mirror when the cavity is blocked. We assume that the spatial mode originating from the cavity is not coupled with a higher efficiency to the fiber than the reflected mode that we use for comparison.
	We detect the reflected pulses and the AFC echoes with a single photon avalanche detector (PerkinElmer SPCM-AQR-16-FC, \qty{25}{\Hz} dark counts).
	Note that we do not need to account for losses along the path, since the echo and the reflected input are transmitted through the same path. The efficiency can be understood as a device efficiency where the cavity around the vacuum chamber and \PrYSO - crystal constitute the quantum memory. 
	
	\section{Results}
	
	\subsection{High-efficiency storage of weak coherent states}
\begin{figure}[!h]
		\centering{\includegraphics[scale=0.35]{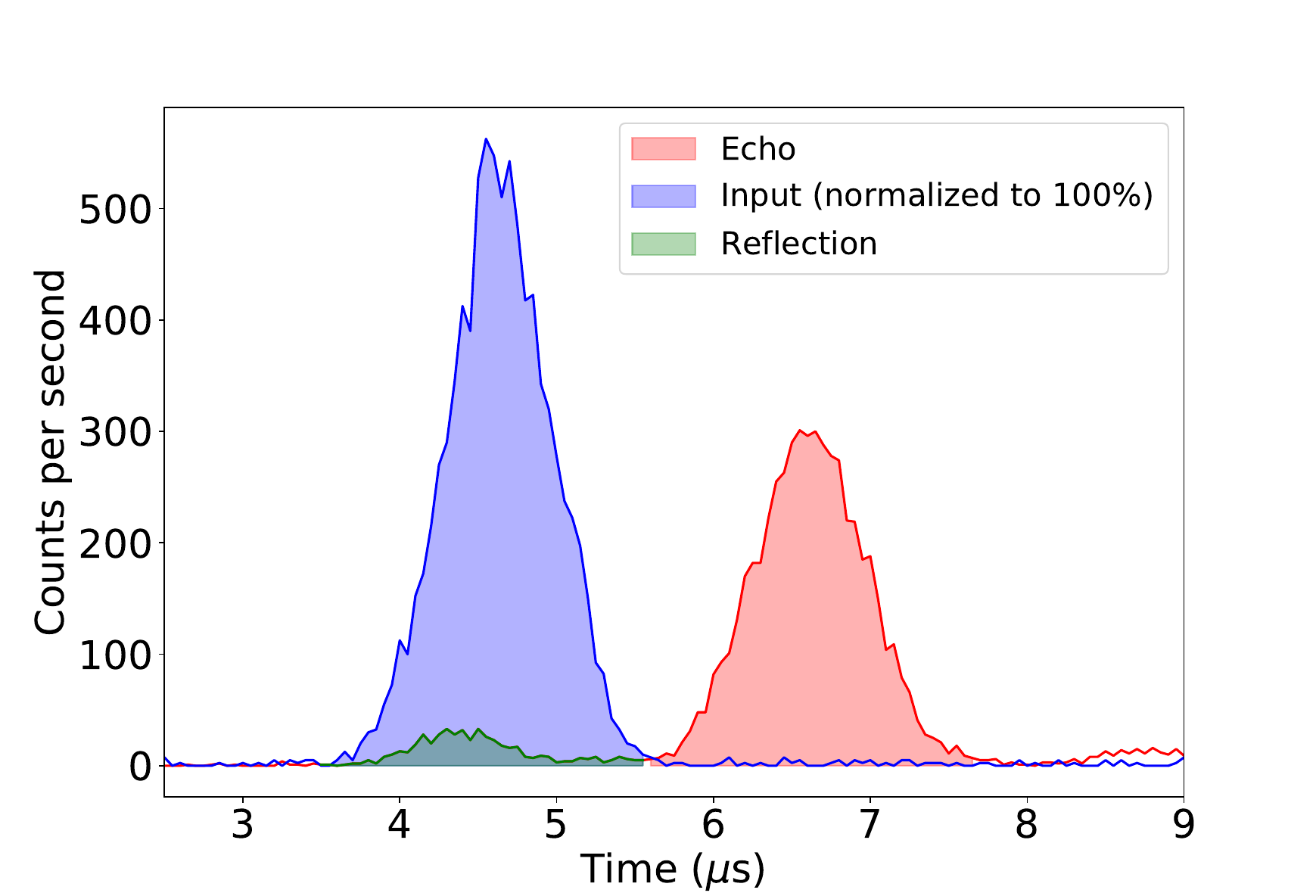}}
		\caption{\raggedright Input pulse and its echo (red pulse) after \qty{2}{\micro\s}. The recorded input { pulse was reflected from the first cavity mirror (\qty{40}{\percent} reflectivity) and} normalized to \qty{100}{\percent} (blue pulse) { of the input signal.} The ratio between the echo and the {input} pulse yields {\maxeff} storage efficiency.  
			The reflection from the cavity ({green} pulse) was \qty{6}{\percent}, indicating that we achieved a good mode-match and a good impedance-match. The detection window for input and echo is \qty{2}{\micro\s}.} 
		\label{fig:im_best_eff_AFC}
	\end{figure}

	We first demonstrate the storage and retrieval of weak coherent pulses with $\mu_{in}$=0.33 for a pre-determined storage time $\tau_{AFC}= \qty{2}{\micro\second}$. Fig \ref{fig:setup_pic} c shows an example of the associated atomic frequency comb, with a measured finesse of 5.8 and an average OD of $\tilde{d}$ =0.4.  Fig.~\ref{fig:im_best_eff_AFC} shows a time histogram with the input pulse and the AFC echo.  
	The storage and retrieval efficiency is $\eta$={\maxeff}, which is to our knowledge the highest AFC efficiency reported up-to-date. 
	The reflected part of the pulse from the cavity is as low as \qty{6}{\percent}, demonstrating a high degree of both mode and impedance matching. 
	The average efficiency measured with ten independent measurements taken over a few days is \qty[uncertainty-mode=separate]{55\pm2}{\percent}.
	For these results, we are sending \num{1000} pulses per second to the cavity, using only the first \qty{12.2}{\milli\s} of the measurement phase.
	The measurement times are ranging between \qty{2}{\min} and \qty{10}{\min}, which ensures a good statistics during all the trials.
	 Taking into account the measured intra-cavity {single-pass } loss of \qty{3}{\percent},{ which is a combination of losses due to imperfect removal of population between the AFC peaks ($\sim \qty{1}{\percent}$) and losses due to scattering and reflection on the surfaces ($\sim \qty{2}{\percent}$),} the finesse of the AFC and the current $\tilde{d}=0.40$, the maximum efficiency is estimated to be \qty{67}{\percent}. {  By reducing the single-pass losses to $\qty{1}{\percent}$ and changing the reflectivity of the second mirror to $\qty{99.5}{\percent}$, the predicted efficiency would be over $\qty{80}{\percent}$. }
 
	\subsection{Storage of weak coherent states for longer storage times}
\begin{figure}[!h]
		\centering{\includegraphics[scale=0.35]{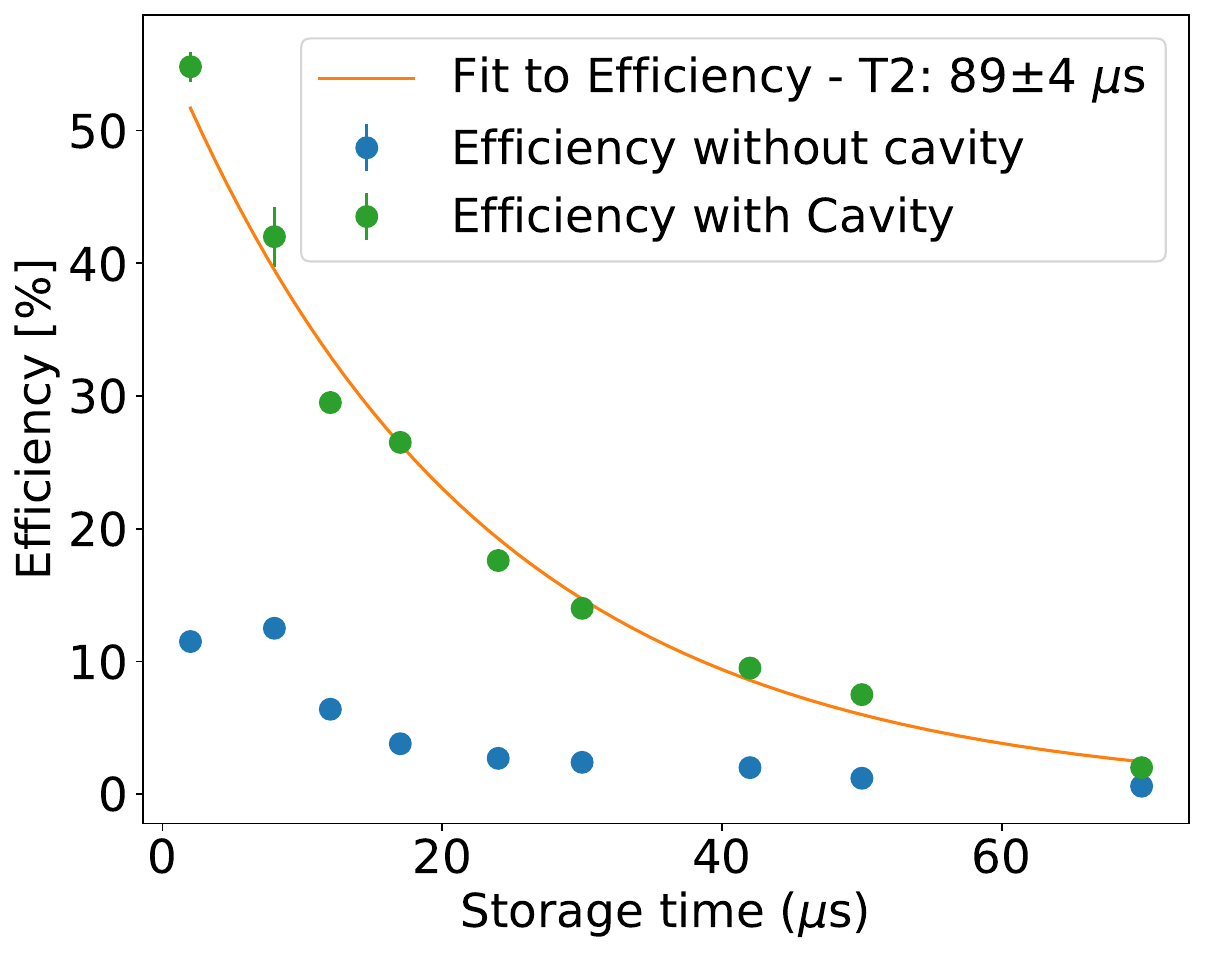}}
		\caption{\raggedright Efficiency versus storage time for 9 different storage times. The blue dots represent the efficiencies with weak coherent pulses, the green ones are the efficiencies in single-pass (without cavity) recorded at the classical level. The orange curve is an exponential fit to the {green} points yielding the effective T$^\text{eff}_2$ of our atomic frequency comb. The combs for the first \num{2} efficiencies are shaped through the hole-burning technique, while the remaining \num{7} combs are prepared through the coherent method~\cite{jobez2016highly}.} 
		\label{fig:eff_vs_storage_time}
	\end{figure}

	We then measured how the efficiency depends on the storage time. For each storage time, we optimized the AFC finesse and height to achieve approximate impedance-matching.	
	Fig.~\ref{fig:eff_vs_storage_time} shows the recorded efficiency as a function of storage time. In order to demonstrate the enhancement, we record both the cavity efficiency  and the single pass efficiency with the same AFC, using coherent pulses with $\mu_{in}$=0.22. Note that the single pass efficiency could still be improved if the OD and finesse of the AFC were optimized for single pass storage. 
	
	The cavity allows us to observe AFC echoes for storage times up to $\tau_{AFC}= \qty{70}{\micro\second} $, where the efficiency was still $\eta$ = \qty{2}{\percent}. By fitting our data with the function $\eta= \eta_{0}\exp\Big(-\frac{4\tau_{AFC}}{T^\text{eff}_2}\Big)$ \cite{jobez2016highly}, where $\eta_{0}$ is the extrapolated efficiency at zero storage time and $\tau$ is the storage time, we can extract T$^\text{eff}_2$=\qty[uncertainty-mode=separate]{89\pm4}{\micro\s}, compatible with previous values reported in free space \cite{lago-rivera2021telecomheralded}.

	\subsection{Bandwidth of the memory}
	Another important parameter is the bandwidth of the memory. As observed in previous experiments with impedance-matched cavities \cite{sabooni2013efficient,sabooni2013cavityenhanced,sabooni2013spectral}, the abrupt change in OD between the inhomogeneously broadened line and the spectral window that is created prior to the AFC preparation leads to strong dispersion and to a dramatic change in the group velocity of light. As such, the free spectral range of the cavity increases therefore, the cavity line width decreases. A measurement of the cavity line width at the center of the spectral window gives \qty{1.16}{\mega\Hz}.
	To assess how this reduction in line width affects the efficiency of spectrally broader pulses, we performed a storage experiment where we change the bandwidth of the input pulses and measure the storage efficiency. The setup for the experiment is the same as the one presented above,  with   \qty{2}{\micro\s} storage time; in this case, we changed the temporal duration of the Gaussian pulses from \qty{120}{\nano\s} up to \qty{1}{\micro\s}, corresponding to pulse bandwidths ranging from \qty{3.7}{\mega\Hz} to \qty{412}{\kilo\Hz}.
	Since the line width of the laser is around \qty{3}{\kilo\Hz} {(FWHM)} and therefore much smaller than these values, we can assume that these pulses are Fourier-limited.
	\begin{figure}[!ht]
		\centering{\includegraphics[scale=0.45]{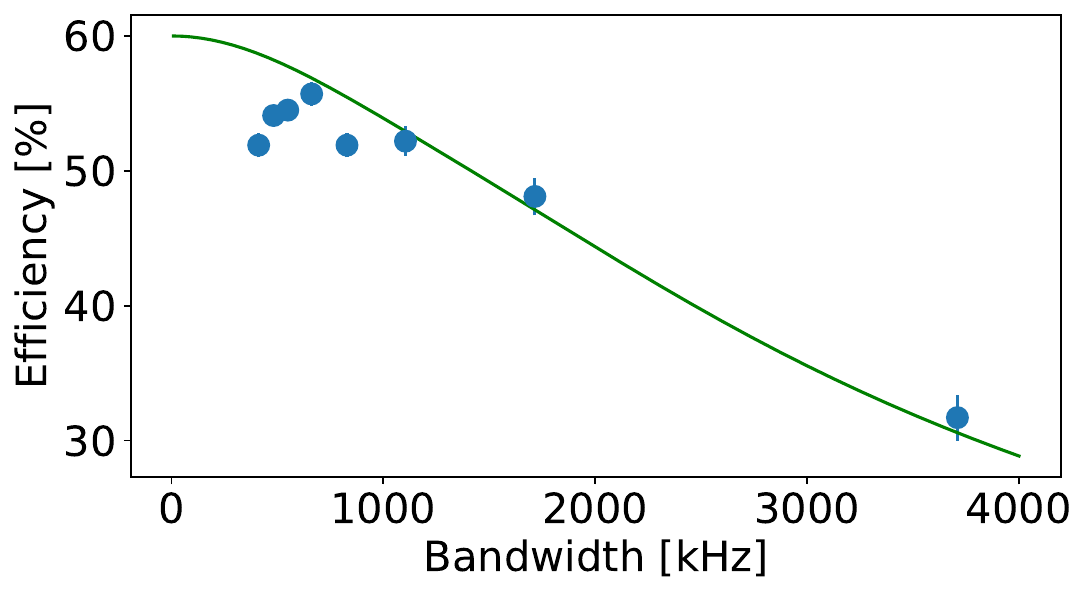}}
		\caption{\raggedright Storage efficiency versus pulse bandwidth {(FWHM)} for Gaussian pulses and \qty{2}{\micro\s} storage time. The input photon number per pulse is 0.32 photons per pulse for the longest pulses and has been decreased for the shorter pulses as the pulse peak power was kept constant. As it is expected, the broader the pulse becomes, the more the storage efficiency decreases, due to the filtering effect of the cavity. The solid line represents the model based on the cavity transfer function which is explained in the main text.  } 
		\label{fig:im_eff_vs_bandwidth}
	\end{figure}
	
	Fig.~\ref{fig:im_eff_vs_bandwidth} shows the results of this experiment. As expected, the efficiency starts dropping as soon as the pulse spectral width exceeded the cavity line width. Despite the efficiency decrease, we can still store a \qty{3.7}{\mega\Hz} broad pulse with \qty{31.7}{\percent} efficiency. Moreover, this measurement tells us that we would be able to store { pulses with a bandwidth of \qty{1.8}{\mega\Hz} (FWHM)} with \qty{48}{\percent} efficiency. This bandwidth corresponds to the one of single photons generated in our cavity enhanced spontaneous parametric down-conversion source~\cite{rielander2014quantum}.

To understand better the decay, we include a simple model in the plot that accounts for the spectral overlap  of the input pulse with the transfer function of the cavity. The pulses are modelled as transform-limited Gaussian pulses with a bandwidth (FWHM) of $2\sqrt{2\ln2}\sigma$ and the overlap  $T(\sigma)$ with the transfer function $H(\omega)$ was determined within the part of the spectrum that contains the AFC and has a width of $\Delta_{\mathrm{AFC}}=\qty{3.5}{\MHz}$. The overlap is given by  
  \begin{equation}
      T(\sigma)= \frac{1}{\sigma \sqrt{2\pi}}\int_{-\Delta_\mathrm{AFC}/2}^{+\Delta_\mathrm{AFC}/2}d\omega |H(\omega)|^2 \cdot \mathrm{e}^{-\frac{\omega^2}{2\sigma^2}}\label{eq:overlapp-integral}. 
  \end{equation}
The transfer function  $H(\omega)$  of a symmetric Fabry-Perot interferometer~\cite{diels2006ultrashort} is given by \begin{equation}
     H(\omega) = \frac{(1-r)\mathrm{e}^{i t_r \omega    }}{1-r\mathrm{e}^{2it_r \omega }}.\label{eq:fabry_perot_transfer_function}
 \end{equation}
 As mentioned above, inside the transparency window without AFC, the line width of the cavity is reduced by a factor 96 to \qty{1.16}{\mega\Hz} compared to the empty cavity. In the model, this means that the round-trip time $t_r$ is  \qty{133}{\micro\second}.  Furthermore, in presence of an AFC which fulfills the impedance-matching condition, the effective round-trip loss is \qty{60}{\percent}, therefore the system is equivalent to a symmetric cavity with $r=\sqrt{0.4}$ (see Eq.~\ref{eq:fabry_perot_transfer_function}). 
 The curve $\eta(\sigma)$, which is displayed in Fig.~\ref{fig:im_eff_vs_bandwidth}, is given by $\eta(\sigma)=\eta_0T(\sigma)$ with $\eta_0=\qty{60}{\percent}$.

  This model fits the data well in the regime of short pulses, where the efficiency is mainly limited by the mismatch between the bandwidth of the pulse and  the cavity. For smaller line width below 1 MHz, the agreement is less good. This may be due to the fact that in this regime the photon spectrum covers fewer AFC teeth and that the efficiency is more sensitive to the quality of the preparation of the comb, which was not optimized for each point in this measurement. This would also explain why the efficiencies achieved are slightly lower than in Fig.~\ref{fig:im_best_eff_AFC}. 

	\subsection{Storage of photonic time-bin qubits}
	Finally, we show that our fixed-storage-time quantum memory can efficiently and faithfully store photonic quantum bits. To this aim, we generate, store,  retrieve and analyze weak coherent time-bin qubits of the form $\ket{\psi}=1/\sqrt{2} ( \ket{e} + e^{i\delta} \ket{l}$), where $\ket{e}$ ($\ket{l}$) is the early (late) component. 
	
	The qubit is \qty{1}{\micro\s}-long, with each Gaussian pulse being \qty{510}{\nano\s} long (FWHM), and contains on average $\mu_{in}$ =0.25 photons per qubit. 
	
	The memory storage time was \qty{2}{\micro\s}. The qubit echo is emitted in the reflected path, coupled to a SM fiber and routed towards the analyzing crystal. Inside this  crystal, we prepared another AFC that we used as an unbalanced Mach-Zehnder inteferometer: we tune the OD of the inteferometric comb in such a way that the probabilities of transmitting the input without absorption (short path) or absorbing and re-emitting them as echoes (long path)~\cite{jobez2015stockage}, are equal. 
	The long path delay, i.e. the AFC storage time, of this interferometer corresponds to the time between the two input time-bins (\qty{1}{\micro\s}): in this way,  the process corresponding to the early time bin taking the long path in the analyzer and the late time-bin taking the short path in the analyzer will be superposed. This enables interference between the two processes and from the visibility of the interference fringe we can assess the fidelity of the equatorial qubit storage. The phase $\Delta \Phi$ of the interferometer can be scanned by shifting the AFC spectral position with respect to the central frequency of the qubit \cite{afzelius2009multimode,jobez2015stockage}. A schematic of this qubit storage and interference is included in Fig.~\ref{fig:setup_pic}. 
	
	In a first step, we prepare a transparent spectral window inside the filter crystal, to assess the storage efficiency of our qubit memory. We recorded two sets of data, exhibiting \qty{51}{\percent} efficiency.
	We attribute in part the lower efficiency compared to the previous single pulse data to the larger bandwidth of the qubit pulses. Applying the model described in the previous section yields an efficiency of \qty{56.3}{\percent}. This is equivalent to the efficiency the model predicts for a Gaussian pulse with a bandwidth of \qty{730}{\kilo\Hz}. The further reduction may be due to comb preparation imperfections or slightly higher intracavity loss for this measurement. Nevertheless, this efficiency is to our knowledge the highest demonstrated for  qubit storage in a solid-state quantum memory.  
	
\begin{figure}[!htbp]
		\centering{\includegraphics[scale=0.25]{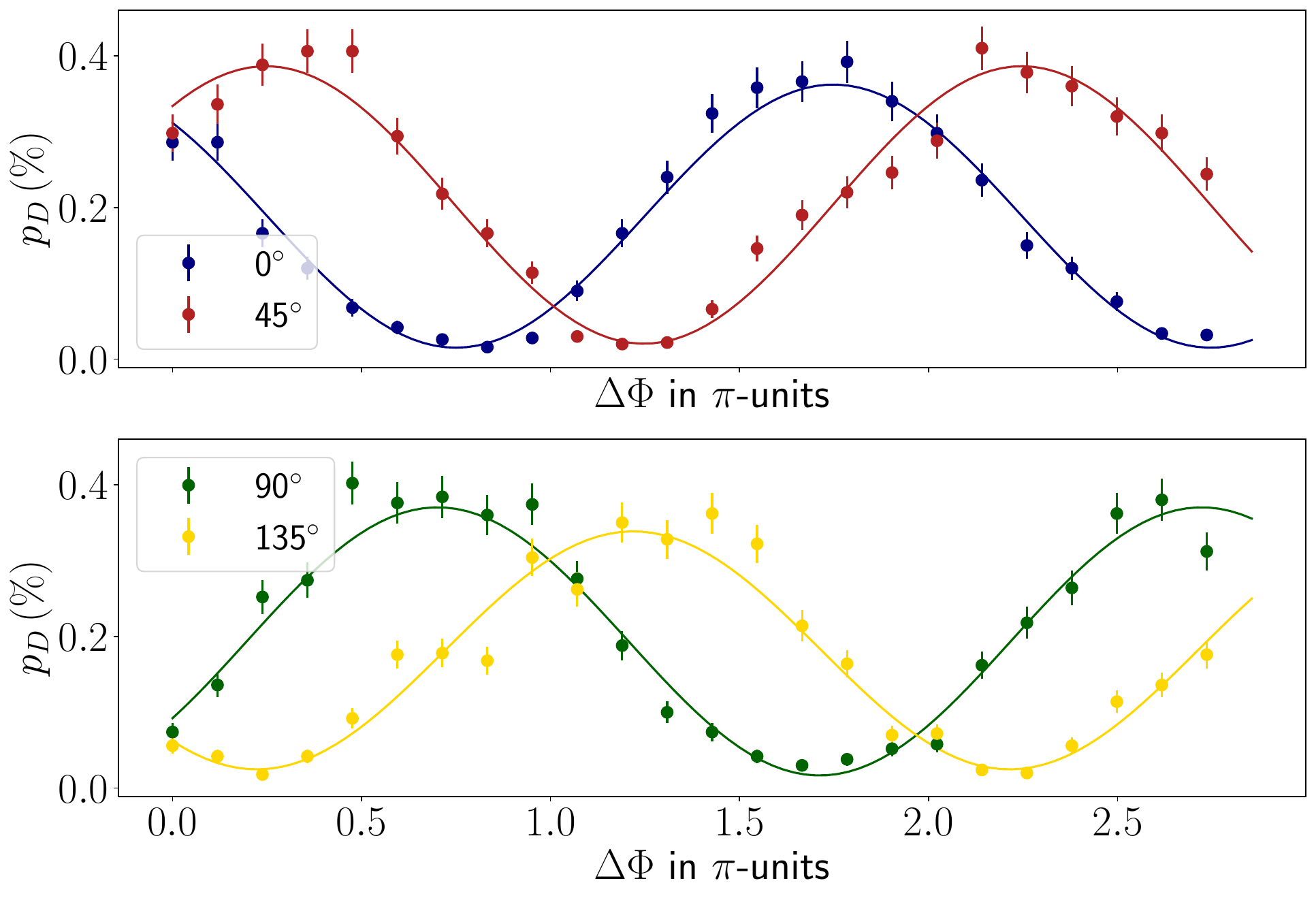}}
		\caption{\raggedright Interference fringes for different phases of the qubit. The plots show the detection probability per trial (p$_\text{D}$) versus the interferometer phase change, expressed in units of $\pi$. All these measurements were recorded with \num{2000} pulses per second. The detection window was \qty{1}{\micro\s}.} 
		\label{fig:im_qubit_interf}
	\end{figure}

	We then prepared the AFC inside the analyzing crystal and recorded the detection probability as a function of the analyzer phase $\Delta \Phi$ for various input qubit states ($\ket{\pm}=1/\sqrt{2} (\ket{e} \pm \ket{l}$) and $\ket{L/R}=1/\sqrt{2} (\ket{e} \pm i\ket{l})$. The interference fringes are shown in Fig.~\ref{fig:im_qubit_interf} and the fitted visibilities are reported in Table~\ref{tab:qubit_vis}. The final average value of $V_\text{coh}\,$=$\,$\qty[uncertainty-mode=separate]{89.9(3.9)}{\percent}, calculated by averaging the  measurements corresponding to the same basis with equal weights. From this visbility, we deduce an  average fidelity for the superposition states of $F_\text{coh}=\frac{1+V_\text{avg}}{2}$ =\qty[uncertainty-mode=separate]{95(2)}{\percent} \cite{marcikic2003longdistance}. 
	
	\begin{table}[]
		\centering
		\begin{tabular}{ |p{1.15cm}|p{1.75cm}|p{1.75cm}|p{1.75cm}|p{1.75cm}| }
			\hline
			Phases [deg]& \qty{0}{}&\qty{45}{}&\qty{90}{}&\qty{135}{}\\
			\hline
			V [\%]& 91.9$\pm$3.3 & 89.9$\pm$3.4 & 91.4$\pm$3.9 &86.4$\pm$4.9 \\
			\hline
		\end{tabular}
		
		\caption{\raggedright Fringe visibility for different values of the phase between qubit components. \label{tab:qubit_vis}}
	\end{table}
	
	To complete the assessment of the storage fidelity, we also measured the states on the poles of the Bloch sphere, $\ket{e}$ and $\ket{l}$. We find a fidelity of F$_\text{pole}\,$=$\,\qty[uncertainty-mode=separate]{94.6(0.3)}{\percent}$. Finally, the fidelity averaged over the Bloch sphere is F$_\text{total}=\frac{2}{3}\text{F}_\text{coh}+\frac{1}{3}\text{F}_\text{comp}\,$=$\,\qty[uncertainty-mode=separate]{94.8(1.4)}{\percent}$.
	A similar analysis  involving only qubit preparation and analysis, but without the quantum memory leads to a qubit fidelity of F$_\text{interf}\,$=$\,$\qty[uncertainty-mode=separate]{94.2\pm1.0}{\percent}.  Subtracting the background counts which are mainly due to dark counts of the detector, would increase the fidelity of this measurement from \qty{94.2}{\percent} to \qty{95.7}{\percent}. Another effect limiting the fidelity is that the two time-bins of the qubit are very close together and partially overlap. Reducing the size of the detection window would mitigate this, but we fixed the size of the detection window in which we quantified the interference to the same size of \qty{1}{\micro\second} per time-bin that we used to quantify the efficiency of the qubit storage. Correcting the fidelity of the creation- and analysis system for the non-zero overlap, would yield a fidelity of \qty{98.3}{\percent}. We attribute the remaining \qty{1.7}{\percent}, to imperfections in the measurement device, for example a non-balanced splitting ratio, and imperfections in the preparation, due e.g. to the finite laser line width of 3 kHz.    
 All this shows that the storage fidelity that we measured, was limited by the creation and analysis system and
 therefore we can estimate the fidelity {$\text{F}_\text{memory}$} of the memory alone {to be compatible with \qty{100}{\percent} within the statistical uncertainty of \qty{2}{\percent}   $(\text{F}_\text{memory}=\text{F}_\text{total}/\text{F}_\text{interf}=\qty{100}{\percent}_{\qty{-2}{\percent}}^{\qty[retain-explicit-plus=true]{+0}{\percent}})$.}
	
	The measured fidelity is above the threshold of 2/3 for measure-and-prepare strategies for single photon input states. However, since we use weak coherent states, the threshold fidelity is increased when we include the Poissonian statistics of the input state \cite{Specht2011, gundogan2012quantum}. For a photon number around \num{0.25} and a memory efficiency of \qty{51}{\percent} (for qubits), we find a threshold of \qty{75}{\percent}. Our measured fidelity is much higher than this bound, showing that our device performs genuine quantum storage.

	\bigskip 
	
	\section{Conclusions and Outlook}
	To summarize, we have demonstrated a cavity-enhanced atomic frequency comb memory at the single photon level. Using an impedance-matched cavity configuration, we showed the storage of weak coherent pulse with efficiencies up to \qty{62}{\percent}  for a fixed storage time of \qty{2}{\micro\second}, which represents the most efficient AFC memory at the single photon level to date. In addition we demonstrated the storage of photonic time-bin qubits with an efficiency of \qty{51}{\percent}, and a fidelity of \qty{95}{\percent} for the retrieved, only limited by the qubit preparation and analysis system. This represents the most efficient storage of photonic qubits in solid-state quantum memories up-to-date. 
	
	The efficiency enhancement due to the cavity was observed for AFC storage times up to \qty{70}{\micro\second}. 
	
	The maximum efficiency achievable in our current setup was limited by intra-cavity loss, by the limited finesse of the AFC, and by the non-unity reflectivity of the second mirror. Intra-cavity loss could be reduced by installing the cavity inside the vacuum chamber of the cryostat, or by coating the crystal facets, to avoid losses at the window surfaces. If loss can be reduced, the effective $\tilde{d}$ could also be reduced which would allow us to reach even higher efficiencies. For a loss of \qty{1}{\percent} and comb finesse of 10, the maximum efficiency would reach \qty{91}{\percent}. Another limitation of the current configuration is the limited bandwidth of the memory due to the slow light effect. This effect could be alleviated by reducing the absorption in the AFC, e.g. by detuning the light away from the center of the inhomogeneous absorption line or by using a shorter crystal. In this work, light was stored for pre-determined AFC storage times in the excited state of the Pr ions. While this could be already useful for some quantum repeater protocols \cite{sinclair2014spectral}, on-demand read-out could also be implemented by sending strong control pulses to store the light in the spin states of Pr ions~\cite{seri2017quantum}.     
	
	Finally, a Faraday rotator could be used in combination with a PBS to replace the beam-splitter and double the detected signal. These improvements would open the door to storing single photons  (e.g. heralded single photons from a cavity-enhanced spontaneous parametric down-conversion source~\cite{rielander2014quantum}) with high efficiencies in solid-state quantum memories, which would represent a critical resource for the realization of functional quantum repeaters.

	\bigskip
	
	\medskip

	\begin{backmatter}
		\bmsection{Funding}
		This project received funding from the European Union research and innovation program within the Flagship on Quantum Technologies through Horizon 2020 grant 820445 (QIA) and  Horizon Europe project QIA-Phase 1 under grant agreement no. 101102140 , from the Government of Spain (PID2019-106850RB-I00; Severo Ochoa CEX2019-000910-S; BES-2017-082464), from the MCIN with funding from European Union NextGenerationEU PRTR (PRTR-C17.I1; MCIN/AEI/10.13039/501100011033: PLEC2021-007669 QNetworks), from Fundaci\'o Cellex, Fundaci\'o Mir-Puig, and from Generalitat de Catalunya (CERCA, AGAUR).

		\bmsection{Data availability} The experimental data underlying the results presented in this paper may be obtained from the authors upon reasonable request.
		
		\bmsection{Disclosures}
		The authors declare no conflicts of interest.

		

	\end{backmatter}
	
\end{document}